\documentclass[hyper,12pt,letterpaper]{JHEP3}
\usepackage{amsmath,amssymb,epsfig}

\font\cmss=cmss12

\newcommand{\ra}{\rightarrow}

\newcommand{\bc}{\begin{center}}
\newcommand{\ec}{\end{center}}
\newcommand{\ba}{\begin{array}}
\newcommand{\ea}{\end{array}}
\newcommand{\beq}{\begin{equation}}
\newcommand{\eeq}{\end{equation}}
\newcommand{\bea}{\begin{eqnarray}}
\newcommand{\eea}{\end{eqnarray}}
\newcommand{\bmx}{\begin{pmatrix}}
\newcommand{\emx}{\end{pmatrix}}
\newcommand{\nn}{\nonumber}

\newcommand{\half}{\frac{1}{2}}
\newcommand{\tr}{{\rm tr\,}}

\newcommand{\tC}{{\tilde C}}

\newcommand{\eref}[1]{Eq.~(\ref{#1})}

\newcommand{\vQ}{{\vec{Q}}}
\newcommand{\vP}{{\vec{P}}}
\newcommand{\vQR}{{\vec{Q}_R}}
\newcommand{\vPR}{{\vec{P}_R}}

\newcommand{\bF}{{\bf F}}

\newcommand{\mbps}{M_{\rm BPS}}

\def\IB{\relax{\rm I\kern-.18em B}}
\def\IC{{\relax\hbox{\kern.3em{\cmss I}$\kern-.4em{\rm C}$}}}
\def\ID{\relax{\rm I\kern-.18em D}}
\def\IE{\relax{\rm I\kern-.18em E}}
\def\IF{\relax{\rm I\kern-.18em F}}
\def\II{\relax{\rm I\kern-.18em I}}
\def\IZ{\relax{\sf Z\kern-.35em Z}}
\def\Id{\relax{1\kern-.32em 1}}
\def\IG{\relax\hbox{$\inbar\kern-.3em{\rm G}$}}
\def\IR{\relax{\rm I\kern-.18em R}}

\newcommand\ncr[2]{\begin{pmatrix}{#1}\\{#2}\end{pmatrix}}
\newcommand\pmat[4]{\begin{pmatrix}{#1}~ &~{#2}\\{#3}~&~{#4}\end{pmatrix}}

\normalsize

\title{Dyon Death Eaters} \author{Anindya Mukherjee\,\footnote{Email:
    anindya\_m@theory.tifr.res.in}\,, Sunil Mukhi\,\footnote{Email:
    mukhi@tifr.res.in}\,, and Rahul Nigam\,\footnote{Email:
    rahulnig@theory.tifr.res.in}\\ \it Tata Institute of Fundamental
  Research,\\ \it Homi Bhabha Rd, Mumbai 400 005, India} \abstract{We
  study general two-body decays of primitive and non-primitive
  $\frac14$-BPS dyons in four-dimensional type IIB string
  compactifications. We find a ``master equation'' for marginal
  stability that generalises the curve found by Sen for $\half$-BPS
  decay, and analyse this equation in a variety of cases including
  decays to $\frac14$-BPS products. For $\half$-BPS decays, an
  interesting and useful relation is exhibited between walls of
  marginal stability and the mathematics of Farey sequences and Ford
  circles. We exhibit an example in which two curves of marginal
  stability intersect in the interior of moduli space.}

\preprint{ arXiv:0707.3035 [hep-th] \\ TIFR/TH/07-14}


\begin{document}

\section{Introduction}
\label{Introduction}

In the last couple of years there has been renewed interest in the
properties of dyonic black holes in four dimensions, particularly
those associated to ${\cal N}=4$ compactifications (type II strings on
$K3\times T^2$ or heterotic/type I strings on $T^6$, as well as
supersymmetry-preserving orbifolds of these systems)
\cite{Shih:2005uc,Gaiotto:2005hc,Jatkar:2005bh,David:2006ji,%
  David:2006yn,David:2006ru, David:2006ud,Dabholkar:2006bj,%
  Sen:2007vb,Dabholkar:2007vk,Banerjee:2007ub,Sen:2007pg,%
Cheng:2007ch,Sen:2007nz}.
A key advance has been a better understanding of a classic degeneracy
formula due to Dijkgraaf, Verlinde and
Verlinde\cite{Dijkgraaf:1996it}. Among other things, the
generalisation of this formula to CHL orbifolds and the origin of a
genus-2 modular form have been illuminated in many of these works.

Recent work has focused on the issue of marginal stability of these
dyons. Curves of marginal stability for specific decays have been
obtained\cite{Sen:2007vb}, the impact of such decays on the degeneracy
formula has been
studied\cite{Sen:2007vb,Dabholkar:2007vk,Cheng:2007ch} and the decays
across such walls have been identified with the disappearance of
two-centred black holes from the
spectrum\cite{Sen:2007pg,Cheng:2007ch}, following previous ideas in
the ${\cal N}=2$ context\cite{Denef:2007vg}.  A formula has been
proposed in \cite{Cheng:2007ch} to count the ``immortal'' dyons which
exist everywhere. And very recently, Sen has considered the case of
primitive dyons decaying into $\frac14$-BPS states\cite{Sen:2007nz}
and demonstrated that this takes place only on surfaces of codimension
2 in moduli space.

Clearly there is much more to be learned about this system. Among
various interesting questions is a complete understanding of the
possible marginal decays of $\frac14$-BPS dyons, the impact of such
decays on the degeneracy counting function, the role of multi-centred
black holes in the decays, and the relevance of ``non-primitive''
dyons (which are related to Riemann surfaces of genus $g>2$) to the
counting problem.

In the present work we take a step towards resolving the first
problem. We consider the most general decay of a $\frac14$-BPS dyon
into two decay products, each one of which can be either $\half$- or
$\frac14$-BPS. We find a necessary condition for marginal two-body
decays and study the resulting equation in a variety of cases. It
turns out that some solutions of our equation are ``spurious'' in the
sense that they describe an inverse decay process rather than the
forward decay. This puts constraints on the possible decay products
which are identical to those found in \cite{Sen:2007nz}. We are also
able to reproduce some of the results in Refs.\cite{Sen:2007vb} as a
special case, as well as generalise them to the case of
``non-primitive'' dyons. On the way we will see that a known
mathematical construction, that of Farey sequences and Ford circles,
bears a remarkably close relation to the circles of marginal stability
in Ref.\cite{Sen:2007vb} and helps us understand the properties of
these circles.

\section{The system}

We consider type IIB string theory compactified on $K3\times T^2$. The
resulting four-dimensional system has 28 $U(1)$ gauge fields and their
electric-magnetic duals. Therefore we can have dyons of charge
$(\vQ,\vP)$ where the first entry is a 28-component vector denoting
electric charge under these gauge fields and the second denotes the
magnetic charge. The dyons will be $\half$-BPS if the vectors
$\vQ,\vP$ are parallel, and $\frac14$-BPS otherwise. 

Note that a modular transformation of the 2-torus $T^2$ that changes
its modular parameter as:
\beq
\tau\to \frac{a\tau + b}{c\tau +d}
\eeq
with 
\beq
\pmat{a}{b}{c}{d}\in PSL(2,Z)
\eeq
sends the dyon charges to:
\beq
\label{dyonmod}
\ncr{\vQ}{\vP} \to \ncr{a\vQ + b\vP}{c\vQ + d\vP}
\eeq

We are interested in the marginal decays of these $\frac14$-BPS
dyons. The stability or otherwise is dictated by the charges carried
by the dyons as well as the values of the moduli of $K3\times
T^2$. These are encoded as follows. Define the matrix:
\beq
L\equiv {\rm diag}(1^6;(-1)^{22})
\eeq
In 4 dimensions there are, first of all, 132 moduli that can be
assembled into a matrix $M$ that is symmetric and orthogonal with
respect to the $L$ metric:
\beq
M^T=M,\quad M^T LM = L
\eeq
The relevant inner product for an electric charge vector, which we will call
$Q^2$ or $\vQ\cdot\vQ$, is\footnote{Because our focus is on
  microstates, our inner products are always defined with respect to
  the moduli at infinity, so this notation should not cause confusion.}:
\beq
Q^2 \equiv \vQ^T(M+L) \, \vQ
\eeq
Correspondingly we have:
\bea
P^2 &\equiv& \vP^T(M+L)\,\vP \nn\\
P\cdot Q &\equiv& \vP^T(M+L)\,\vQ
\eea
We will also make use of the quantities $\vQ_R,\vP_R$ defined such
that 
\beq 
Q_R^2\equiv\vQ_R^T\vQ_R=\vQ^T(M+L) \, \vQ 
\eeq 
and similarly for
the other inner products (for details see for example
\cite{Cheng:2007ch,Sen:2007nz}).

In addition to the moduli appearing in $M$, there is the modular
parameter of the 4-5 torus:
\beq
\tau = \tau_1+i\tau_2
\eeq
The BPS mass formula for general $\frac14$-BPS dyons is
\cite{Cvetic:1995uj,Duff:1995sm}:
\beq
\label{genbpsmass}
\mbps(\vQ,\vP)^2 =
\frac{1}{\sqrt\tau_2}(\vQ-{\bar\tau}\vP)\cdot(\vQ-{\tau}\vP) +
2\sqrt{\tau_2}\sqrt{\Delta(\vQ,\vP)} \eeq
where
\beq
\label{Deltadef}
\Delta(\vQ,\vP) \equiv Q^2 P^2 - (P\cdot Q)^2
\eeq

Before going on, it is useful to transform the dyon charges to bring
them into a standard form. Consider the electric and magnetic charge
vectors $\vQ,\vP$ of the dyon and define\cite{Dabholkar:2007vk}:
\beq
I(\vQ,\vP) \equiv gcd(\vQ\wedge\vP) = gcd(Q^iP^j-Q^jP^i),~\hbox{all
  $i,j$}
\eeq
If for a given dyon we find that $I(\vQ,\vP)>1$, we first perform an
$SL(2,Z)$ transformation as in \eref{dyonmod}. Using some properties
of finitely generated algebras (see for example
Ref.\cite{Langalgebra}, Chapter I, Section 8), we can always find such
a transformation\footnote{We are grateful to Nitin Nitsure for helpful
  discussions on this point.} that yields new dyon charges of the form
$(m\vQ',n\vP')$ for some positive integers $m,n$ and some new vectors
$\vQ',\vP'$ such that $I(\vQ',\vP')=1$. Under this transformation
$I(\vQ,\vP)$ remains invariant, so $m,n$ must be such that
$I(\vQ,\vP)=mn$. If the $m,n$ so obtained are not co-prime then the
dyon with those $m,n$ will be marginally unstable at all points of
moduli space. This does not mean a bound state does not exist, but
that determining its existence is more subtle and requires actually
quantising the system. Therefore we will restrict ourselves to
the case where $m,n$ are co-prime.

To summarise, in what follows we assume that our dyons have charge
vectors $(m\vQ,n\vP)$ with co-prime $m,n$ and with $I(\vQ,\vP)=1$. The
special case $(m,n)=(1,1)$ will be called a {\it primitive} dyon.

\section{Decays into a pair of dyons}

We can now examine the decay of a $\frac14$-BPS dyons into two other
dyons. From charge conservation, the most general decay is of the form:
\beq 
\label{gendecay}
\ncr{m\vQ}{n\vP} \to \ncr{\vQ_1}{\vP_1} +\ncr{m\vQ-\vQ_1}{n\vP-\vP_1}
\eeq 
where $\vQ_1,\vP_1$ are arbitrary vectors in the $(6,22)$-dimensional
integral charge lattice.

From the study of BPS string junctions and
networks\cite{Aharony:1997bh,Dasgupta:1997pu,Sen:1997xi}, we know that
the decay products can be mutually BPS with each other and with the
initial state only if the corresponding charges all lie in a plane
rather than being generic 28-dimensional vectors as above. However,
the properties of the networks are determined in the present context
not by the charge vectors $\vQ,\vP$ but by their projections
$\vQ_R,\vP_R$. Indeed it is only the latter which appear in the BPS
mass formula \eref{genbpsmass} that we will be using. Therefore the
BPS condition requires that the $R$ projections of the final-state charges
are in the same plane as those of the initial-state charges. This
happens automatically in some cases, while in others it requires
adjusting the moduli in $M$ to make this happen.

It follows that we must have the relation:
\beq
\label{gendecayR}
\ncr{m\vQR}{n\vPR} \to \ncr{m_1\vQR+r_1\vPR}{s_1\vQR+n_1\vPR} +
\ncr{m_2\vQR +r_2\vPR}{s_2\vQR+n_2\vPR} 
\eeq 
where the coefficients $m_i,n_i,r_i,s_i$ satisfy:
\beq
m_1+m_2=m,\quad n_1+n_2=n,\quad r_1+r_2=s_1+s_2=0
\eeq
We cannot, however, assume that these coefficients are integers since
the above equation refers not to the original vectors in the integral
lattice but to their projections to the $\vQR,\vPR$ plane. 

Without any additional conditions on these coefficients the decay products
will both be $\frac14$-BPS dyons. It is possible to have one or both
of them be $\half$-BPS by suitably constraining the integers, as we
will see shortly.

If $M,M_1,M_2$ denote the BPS masses of the initial state and the two
decay products (for simplicity we henceforth drop the subscript
$BPS$), we can use Eqs.(\ref{genbpsmass}) and (\ref{gendecayR}) to
evaluate the condition on the moduli imposed by the marginality
condition $M=M_1+M_2$. Because of the square root in
\eref{genbpsmass}, this is most easily done by computing a combination
of squared masses that vanishes when the marginality condition is
satisfied.

First, define the angles $\theta$ and $\theta_{12}$ by:
\bea
\label{angledef}
\theta&=&\tan^{-1}\frac{\tau_2}{\tau_1}\nn\\
\theta_{QP}&=& \cos^{-1}\frac{Q_R\cdot P_R}{Q_R P_R}
\eea
where $Q_R\equiv |\vQR|, P_R\equiv |\vPR|$. Geometrically, $\theta$ is
the opening angle of the torus while $\theta_{QP}$ is the angle
between the projected electric and magnetic charge vectors (which
coincides with the angle appearing in the string junction description
of the dyon). We also define a ``cross-product'' between the integers
$m_1,n_1,m_2,n_2$ as:
\beq
m\wedge n = m_1 n_2 - m_2 n_1
\eeq

Let us now find the condition that, at some point(s) in moduli space,
the decay \eref{gendecayR} becomes marginal: $M = M_1 + M_2$. The BPS
formula \eref{genbpsmass} involves a square root on the RHS and
another square root to extract $M$ from $M^2$. The simplest
square-root-free expression that vanishes when the marginality
condition is satisfied is the combination:
\bea 
\label{marginalquartic}
&&M^4 + M_1^4 + M_2^4 -2(M^2M_1^2 +M^2M_2^2 +M_1^2M_2^2 )\nn\\
&&\qquad= (M-M_1-M_2)(M+M_1+M_2)(M-M_1+M_2)(M+M_1-M_2)\qquad\qquad
\eea
Now we require this expression to vanish. However, subsequently we
must check that on the vanishing curve, it is really the first factor
of the RHS of \eref{marginalquartic} that vanishes rather than any of
the other factors. Notice that the second factor never vanishes (since
all the $M$'s are positive), while vanishing of the third or fourth
factor corresponds to the inverse decays $M_1=M+M_2$ and
$M_2=M+M_1$. When we turn to a detailed analysis of marginal decay
processes, it will be necessary to rule out these inverse decays
before concluding that we are dealing with the correct decay
mode. Only in the case where both the final products are $\half$-BPS,
this check becomes unnecessary because the reverse process is
forbidden: a $\half$-BPS state cannot decay into a $\frac14$-BPS
state.

Now we use the BPS mass formula \eref{genbpsmass}, the formula for the
decay process \eref{gendecayR}, and and the definitions of the angles
in \eref{angledef}, to find after a tedious calculation that:
\bea M^4 &+& M_1^4 + M_2^4 -2(M^2M_1^2 +M^2M_2^2 +M_1^2M_2^2 ) =
-4\tau_2^2\Bigg[Q_RP_R\frac{\sin(\theta+\theta_{QP})}{\sin\theta}\,m\wedge
  n\nn\\ && +~ r_1 P_R\left(mQ_R\,\frac{\sin\theta_{QP}}{|\tau|\sin\theta}
  + nP_R\right) -s_1Q_R\left(nP_R\,
  \frac{|\tau|\sin\theta_{QP}}{\sin\theta} + mQ_R\right)\Bigg]^2 \eea
Vanishing of the RHS is a necessary condition for marginal stability.

This condition can be usefully rewritten by eliminating the angles
$\theta,\theta_{QP}$ and reverting to $\tau_1,\tau_2$ coordinates for
the modular parameter of the torus. It is convenient to introduce the
following quantity depending on charges of the initial and final
states as well as the moduli:
\beq
E\equiv \frac{1}{\sqrt\Delta}\left(ms_1\, Q_R^2 -nr_1\, P_R^2 - (m\wedge n)Q_R\cdot
P_R\right)
\eeq
Then we find that the equation for marginal stability is:
\beq
\label{cms}
\left(\tau_1-\frac{m\wedge n}{2ns_1}\right)^2
+ \left(\tau_2 + \frac{E}{2ns_1}\right)^2
= \frac{1}{4n^2s_1^2}\Big((m\wedge n)^2 + 4mnr_1s_1 + E^2\Big)
\eeq
This is the ``master equation'' governing all two-body decays of
$\frac14$-BPS states in this theory. However we will need careful
analysis to see when the equation does actually describe such a decay
and what type of decay it describes.

Note first of all that the equation is invariant under the
transformation:
\beq 
r_1\to r_2=-r_1,\quad s_1\to s_2=-s_1,\quad m_1\to
m_2=m-m_1,\quad n_1\to n_2=n-n_1
\eeq
under which $m\wedge n$ and $E$ both change sign. This corresponds to
interchange of the two decay products.

If the RHS of \eref{cms} can be shown to be positive definite, this
will be a circle in the torus moduli space with centre at:
\beq
(\tau_1,\tau_2) = \left(\frac{m\wedge n}{2ns_1}, -\frac{E}{2ns_1}\right)
\eeq
and radius 
\beq
\frac{1}{2ns_1}\sqrt{(m\wedge n)^2 + 4mnr_1s_1 + E^2}
\eeq

Because there is no restriction on the signs of $r,s$, it may appear
that the RHS of \eref{cms} is not positive definite. However, after a
little computation we are able to rewrite it as:
\bea 
(m\wedge n)^2 + 4mnr_1s_1 + E^2 &=& \frac{1}{\Delta}\Big(
\left[(m\wedge n)Q_RP_R -
  (ms_1\,Q_R^2-nr_1\,P_R^2)\cos\theta_{QP}\right]^2\nn\\
&&+(ms_1\,Q_R^2 + nr_1\,P_R^2)^2\sin^2\theta_{QP} \Big) 
\eea
which is a sum of squares. Therefore the equation does indeed describe
a nontrivial circle in every case.

The next step is to check whether this circle intersects the upper
half-plane. There are two cases. If $\frac{E}{s_1}>0$ then the centre
of the circle is in the lower half plane. The circle will then
intersect the upper half plane only if it intersects the real axis,
which happens if:
\beq
(m\wedge n)^2 + 4mnr_1s_1>0
\eeq
It is easy to see that:
\beq
(m\wedge n)^2 + 4mnr_1s_1 = \tr \bF^2 - 2\det \bF
\eeq
where
\beq \bF =
\pmat{nm_1}{nr_1}{ms_1}{mn_1}=\pmat{n}{0}{0}{m}\pmat{m_1}{r_1}{s_1}{n_1}
\eeq
Now if $\alpha_1,\alpha_2$ are the eigenvalues of $\bF$ then:
\beq
\tr \bF^2 - 2\det \bF = (\alpha_1-\alpha_2)^2
\eeq
This is positive if $\alpha_1,\alpha_2$ are both real, and negative if
they are complex conjugate pairs. Therefore when $\frac{E}{s_1}$ is
positive, only decays for which the eigenvalues of $\bF$ are real can
produce genuine curves of marginal stability in the upper half plane
of $\tau$-space.

If $\frac{E}{s_1}<0$ then the circle has its centre in the upper half
plane, and therefore always has a finite region in the upper
half-plane.

\section{Analysis of the marginal stability curves: $\half$-BPS decay products}

\subsection{Equations of the curves}

To analyse the equation of marginal stability we have obtained, let us
first consider the special case when both decay products are
$\half$-BPS. This requires that the electric and magnetic charge
vectors of the decay products be proportional.  The equation for the
charges of the decay products \eref{gendecay} can now be written:
\beq 
\label{bpsdecay}
\ncr{m\vQ}{n\vP} \to \ncr{m_1\vQ + r_1\vP}{s_1\vQ+n_1\vP}
+\ncr{m_2\vQ+r_2\vP}{s_2\vQ+n_2\vP} 
\eeq
with $m_i,n_i,r_i,s_i$ satisfying:
\beq
m_1+m_2=m,\quad n_1+n_2=n,\quad r_1+r_2=s_1+s_2=0
\eeq
and where the electric and magnetic (upper and lower) components of
each charge vector are proportional to each other. Note that this is
the equation for the full, rather than projected, charge vector. The
absence of any term out of the plane of $\vQ,\vP$ comes from the fact
that if such a term were present, it would be impossible to make the
electric and magnetic charges proportional in {\it both} decay
products. Because the above equation is for the full charge vectors,
integrality of the charge lattice requires that $m_i,r_i,s_i,n_i$ are
integers. In case all four integers (for each $i$) have a common
factor then the decay will be into three or more final states. Since
we want to focus on two-body decays, we should exclude such cases.

Proportionality of electric and magnetic charges is equivalent to
requiring that the determinant of the associated matrices vanish:
\beq
\label{firstdet}
\det\pmat{m_1}{r_1}{s_1}{n_1}=0
\eeq
and
\beq
\label{seconddet}
\det\pmat{m-m_1}{-r_1}{-s_1}{n-n_1}=0
\eeq

The first of these equations is solved by the substitution:
\beq
\label{firstcond}
\pmat{m_1}{r_1}{s_1}{n_1}=\pmat{ad}{-ab}{cd}{-bc}
\eeq
where $a,b,c,d$ are defined only upto an overall reversal of sign. The
second equation then tells us that
\beq
\label{secondcond}
mn+ bc\,m -ad\,n=0
\eeq

Suppose now that the original dyon was primitive, namely
$(m,n)=(1,1)$.  In this case \eref{secondcond} becomes
\beq 
\label{adbc}
ad-bc=1 
\eeq 
and therefore the decay products are parametrised by a matrix in
$PSL(2,Z)$.  In going to the coefficients $a,b,c,d$, we see that they
are invariant under the scaling $a,b,c,d\to \lambda a, \lambda^{-1}b,
\lambda c, \lambda^{-1} d$ as well as an exchange $a,b,c,d\to
-b,a,-d,c$. These transformations, along with \eref{adbc} can be used
to show that $a,b,c,d$ are unique integers\cite{Sen:2007vb}.

Making the substitutions $(m,n)=(1,1)$ as well as
\eref{firstcond} in the curve of marginal stability \eref{cms}, and
using the $PSL(2,Z)$ property, we find that the curve reduces to:
\beq
\label{sencms}
\left(\tau_1-\frac{ad+bc}{2cd}\right)^2
+ \left(\tau_2 + \frac{E}{2cd}\right)^2
= \frac{1}{4c^2d^2}\left(1 + E^2\right)
\eeq
where
\beq
\label{Edefinition}
E\equiv \frac{1}{\sqrt\Delta}\left(cd\, Q^2 +ab\, P^2 - (ad+bc)Q\cdot
P\right)
\eeq
This is the equation found by Sen in Ref.\cite{Sen:2007vb}.

These curves are circles with centre at $\frac{ad+bc}{2cd}$ and radius
$\frac{\sqrt{1+E^2}}{2cd}$. They intersect the real axis in the pair
of points
\beq
\label{pairofpoints}
\frac{b}{d}, \frac{a}{c}
\eeq
Sen showed that, for primitive dyons, two different curves never
intersect in the upper half plane but can touch on the real axis in
$\tau$-space. This implies that a given primitive $\frac14$-BPS dyon
can at most be marginally unstable to decay into a single definite
pair of $\half$-BPS dyons at a given point in moduli space.

While the fractions $\frac{b}{d}, \frac{a}{c}$ need not in general be
positive or lie between 0 and 1, they can be brought into the form of
positive fractions between 0 and 1 by a modular
transformation. Suppose for example that $\frac{b}{d}$ does not lie
between 0 and 1. Then for some suitable integer $N$, we define
$b'=b-dN$ such that $0< b'\le d$. For the same $N$) we can show that
$a'=a-cN$ satisfies $0<a'\le c$. As a result, $0<
\frac{a'}{c},\frac{b'}{d}\le 1$. The formula for $E$ above is
unchanged under this transformation if we simultaneously re-define
$\vQ\to \vQ-N\vP$, and the curve of marginal stability is
invariant if we also send $\tau_1\to \tau_1+N$.

To complete the discussion of decays into $\half$-BPS final states, we
need to consider the case of dyons that are non-primitive,
i.e. $(m,n)\ne(1,1)$. In this case we can obtain the curve of marginal
stability by starting from \eref{cms} and making the appropriate
substitutions from \eref{firstcond} and \eref{secondcond}.  The
coefficients $ad,ab,cd,bc$ are still integers but they no longer
describe a matrix in $PSL(2,Z)$. Instead they satisfy the condition:
\beq
\label{mncond}
ad\,n-bc\,m=mn
\eeq
Moreover, one can check that $a,b,c,d$ are not unique in this
case. However only the combinations $ad,ab,cd,bc$ actually appear in
the curve of marginal stability, so this curve is unique and can be
written:
\beq
\label{nonprimitivecms}
\left(\tau_1-\frac{nad+mbc}{2ncd}\right)^2
+ \left(\tau_2 + \frac{E}{2ncd}\right)^2
= \frac{1}{4n^2c^2d^2}\left(m^2n^2 + E^2\right)
\eeq
where
\beq
E\equiv \frac{1}{\sqrt\Delta}\left(mcd\, Q^2 +nab\, P^2 - (nad+mbc)Q\cdot
P\right)
\eeq
This is the most general curve of marginal stability for decay into
$\half$-BPS dyons.

Examining the curve we find that it intersects the real axis at the
points $\frac{a}{c}$ and $\frac{mb}{nd}$.  Even though $m,n$ are
co-prime, we cannot be sure that $mb,nd$ are co-prime, so the latter
fraction is not necessarily reduced to lowest terms. We will discuss
the geometry of these curves in a later subsection.

\subsection{Farey fractions and Ford circles}

In this subsection we briefly review some mathematical constructions
that will facilitate the analysis of the $\half$-BPS curves of
marginal stability. In the mathematical literature one encounters the
notion of a Farey sequence $F_n$ (see for example
Ref.\cite{Apostol}). This is the set of all fractions (reduced to
lowest terms) with denominators $\le n$ and taking values in the
interval between 0 and 1, arranged in order of increasing
magnitude. As an example we have:
\beq
F_5=\left\{\frac01,\frac15, \frac14, \frac13, \frac25, \frac12, \frac35,
\frac23, \frac34, \frac45,\frac11\right\}
\eeq
Relevant properties of Farey sequences, for us, are the following
(more details can be found in Ref.\cite{Apostol}). For any pair of
fractions $\frac{b}{d}$ and $\frac{a}{c}$ that appear consecutively
in {\it any} Farey sequence, we have $ad-bc=\pm 1$. We can always
order them so that the sign is positive, therefore $ad-bc=1$. Given
any such pair, a new fraction called the {\it mediant} is given by:
\beq \hbox{mediant} \left(\frac{b}{d},\frac{a}{c}\right) \equiv
\frac{a+b}{c+d} \eeq The mediant lies between the two members of the
original pair and will occur between them in subsequent Farey
sequences. Moreover, if we define
\beq
\label{mediant}
\frac{h}{k} = \frac{a+b}{c+d}
\eeq
then $hd-kb=1=ak-ch$. Thus the fraction $\frac{h}{k}$ will occur after
$\frac{b}{d}$ as well as before $\frac{a}{c}$ in some Farey sequence.

The above construction, which is seen to be related to the structure
of the discrete group $PSL(2,Z)$, can be geometrically visualised in
terms of circles called {\it Ford circles}. These will turn out to be
helpful in understanding the properties of the Sen circles of
\eref{sencms}. For a pair of co-prime integers $a,c$ such that $0\le
\frac{a}{c}\le 1$, the associated Ford circle\cite{Apostol}
$C\left(\frac{a}{c}\right)$ is a circle centred at $(\frac{a}{c},
\frac{1}{2c^2})$ with radius $\frac{1}{2c^2}$. It is tangent to the
horizontal axis at $\frac{a}{c}$, and can be thought of as ``sitting
above'' this fraction. The size of a Ford circle is inversely
proportional to the square of the denominator of the
fraction. Accordingly the largest possible Ford circles, above the
points $\frac{0}{1}$ and $\frac{1}{1}$, have radius $\half$.

\FIGURE{
\label{ford}
\epsfxsize=10cm
\epsfbox{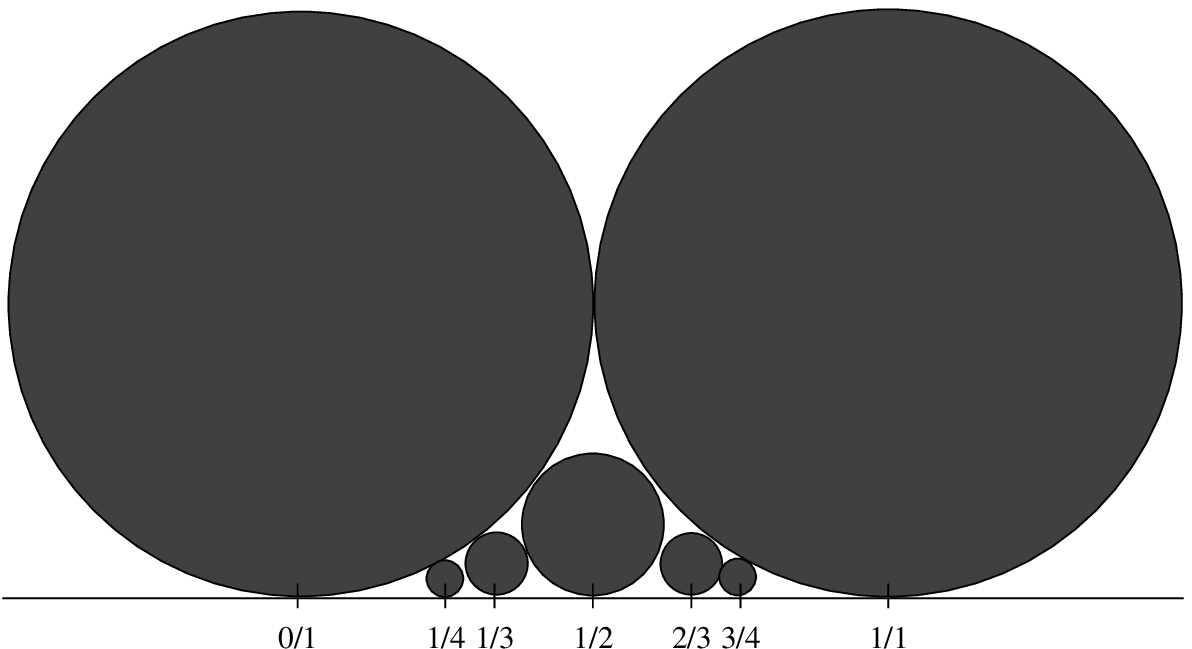}
\caption{The Ford circles associated to $F_4$}}

The key property of Ford circles is that (i) two Ford circles never
intersect, (ii) two Ford circles associated to the fractions
$\frac{b}{d}$ and $\frac{a}{c}$ (without loss of generality we
assume the second fraction to be the larger one) are tangent to each
other if and only if $ad-bc=1$. In terms of Farey sequences,
if two fractions are consecutive in any Farey sequence then they
are associated to a pair of touching Ford circles. Conversely if two Ford
circles touch then their corresponding fractions are consecutive in
some Farey sequence.

Finally we describe a construction that will be closely related to
marginal decays of dyons. For any pair of touching Ford circles
associated to fractions $\frac{b}{d}$ and $\frac{a}{c}$ with
$ad-bc=1$, there is another circle that (for lack of a
better name) we will refer to as the ``dual Ford circle''
$\tC(\frac{b}{d},\frac{a}{c})$ that is centred on the real axis and
passes through the points $\frac{b}{d}$ and $\frac{a}{c}$ on the real
axis. This circle has the property that it also passes through the
point at which the two Ford circles touch\cite{Apostol}.

\subsection{Analysis of the decays: Sen circles and Ford circles}

Now let us return to the decay of a primitive $\frac14$-BPS dyon into two
$\half$-BPS dyons. As we have seen in the previous subsection, the
decay products are defined in terms of a matrix in $PSL(2,Z)$.  This
matrix defines a pair of fractions $\frac{b}{d}$ and $\frac{a}{c}$
with $ad-bc=1$. By the shift $\tau_1\to \tau_1+N$, as in the
discussion below \eref{pairofpoints}, we can make both the fractions
lie between 0 and 1. Now the Ford circles associated to these two
fractions are tangent to each other. The dual Ford circle
$\tC\left(\frac{b}{d},\frac{a}{c}\right)$ has its origin on the real
axis at the midpoint of these two fractions, at
$\frac{ad+bc}{2cd}$. Its radius is given by half the distance between
these two fractions, namely $\frac{1}{2cd}$. Thus the equation of this
dual Ford circle is:
\beq
\label{dfc}
\left(\tau_1-\frac{ad+bc}{2cd}\right)^2
+ \tau_2^2
= \frac{1}{4c^2d^2}
\eeq
Comparing with \eref{sencms}, we see that the dual Ford circle is the
limit of the Sen circle for marginal decays of a primitive
$\frac14$-BPS dyon into two $\half$-BPS dyons, as $E\to 0$. (Recall
that $E$ was defined in \eref{Edefinition}). Conversely, the Sen
circle can be thought of as a deformation of the dual Ford circle with
deformation parameter $E$. For given integers $a,b,c,d$, both circles
are centred at the same value of $\tau_1$ but have their centres
vertically displaced from each other. The radius of the Sen circle is
such that it intersects the real axis in the {\it same} pair of points
as the dual Ford circle. Note that $\frac{E}{cd}$ can be positive or
negative, so the Sen circle can be displaced either downwards or
upwards relative to the dual Ford circle.

This similarity is intriguing and may point to a more profound
relation between Sen circles and Ford circles that we have not yet
uncovered (in particular, it seems plausible that by deforming the K3
moduli one can set $E\to 0$, which would make the two circles actually
coincide). However, already the relation we have exhibited is
sufficient to understand a key property of Sen circles, which is that they
do not intersect in the upper half plane, but only on the real
axis\cite{Sen:2007vb}. 

This can can be seen as follows. Every Sen circle is associated to a
dual Ford circle and thereby to a pair of Ford circles. Consider the
two Sen circles associated to $a,b,c,d$ and $h,p,k,q$ with
$ad-bc=pk-qh=1$. Clearly we have $\frac{b}{d}<\frac{a}{c}$ as well as
$\frac{h}{k}<\frac{p}{q}$.  The two possible orderings of the
fractions are $\frac{b}{d}, \frac{h}{k},\frac{a}{c}, \frac{p}{q}$ and
$\frac{b}{d}, \frac{a}{c}, \frac{h}{k},\frac{p}{q}$.  The first
ordering is ruled out by the Ford circle construction, since it
implies that the Ford circle of the first fraction touches that of the
third one, while the Ford circle of the second fraction touches that
of the fourth one. This contradicts the fact that all the Ford circles
are non-overlapping. Thus only the second ordering is possible, where
we have the fractions $\frac{b}{d}, \frac{a}{c},
\frac{h}{k},\frac{p}{q}$ in increasing order. Let us consider the case
where $\frac{a}{c}=\frac{h}{k}$, so that the Sen circles touch on the
real axis. Clearly the dual Ford circles also touch on the real axis,
which means the three fractions $\frac{b}{d}, \frac{a}{c},\frac{p}{q}$
are consecutive terms in a Farey sequence.

We want to show that the Sen circles in this case do not
intersect in the upper half plane. This imposes a condition on the
slopes of the Sen circles at the real axis. From \eref{sencms} we find
that the slope at the real axis is given by:
\beq
\tan\phi = \pm \frac{1}{E}
\eeq
where the two signs hold for the two intersection points.  Now it is
easy to check that the condition we are seeking is:
\beq
E(a,b,c,d) + E(a,p,c,q)>0
\eeq
This is of course satisfied if both $E$'s are positive, though that is
not in general the case. But even in the general case the condition
above does hold, as we now demonstrate. From the definition of $E$ one
finds that:
\beq
E(a,b,c,d) + E(a,p,c,q)= \frac{1}{\sqrt\Delta}\Big(c(q+d)Q^2 +
a(p+b)P^2 - \left(a(q+d)+ c(p+b)\right)P\cdot Q\Big)
\eeq
Now we use the fact, explained in the discussion around
\eref{mediant}, that if three fractions are consecutive in a Farey
sequence then the middle one is the mediant of the other two. Hence we
have:
\beq
\frac{a}{c}=\frac{p+b}{q+d}
\eeq
from which we get:
\beq
Na=(p+b),\quad Nc=(q+d)
\eeq
for some integer $N\ge 1$. It follows that:
\beq
E(a,b,c,d) + E(a,p,c,q)= \frac{N}{\sqrt\Delta}(c\vQ -a\vP)^2
>0
\eeq
as desired. By similar methods the non-intersecting property of Sen
circles can be proved for the case where $\frac{b}{d}, \frac{a}{c},
\frac{h}{k},\frac{p}{q}$ are all distinct fractions.

\subsection{Analysis of the decays: non-primitive case}

The above discussion was for the case of a primitive dyon as the
initial state. Now let us look at the case where the initial state is
a non-primitive dyon. In this case the Sen circle is replaced by
\eref{nonprimitivecms}, which intersects the real axis at the points
$\frac{a}{c}$ and $\frac{mb}{nd}$. Let us now analyse the condition
\eref{mncond} in some detail. Because $m$ and $n$ are co-prime,
writing this condition as $adn=m(bc + n)$ tells us that $m$ divides
$ad$ and also that $n$ divides $bc$. Therefore we can rewrite
\eref{mncond} as:
\beq
\frac{ad}{m} - \frac{bc}{n} =1
\eeq
where each of the terms on the LHS is an integer. This can only be
realised if, for some (not necessarily prime or unique) factorisation
of $m$ and $n$;
\beq
\label{mnfact}
m=pq, \quad n=kl
\eeq
we have that:
\beq
a'=\frac{a}{p},~~b'=\frac{b}{k},~~c'=\frac{c}{l},~~d'=\frac{d}{q}
\eeq
are all integers. Evidently they satisfy $a'd'-b'c'=1$. Substituting
in the curve of marginal stability for this case,
\eref{nonprimitivecms}, we find: \beq
\left(\tau_1-\frac{p}{l}\frac{a'd'+b'c'}{2c'd'}\right)^2 +
\left(\tau_2 + \frac{p}{l}\frac{E'}{2c'd'}\right)^2 =
\frac{p^2}{4l^2c'^2d'^2}\left(1 + E'^2\right) \eeq where
\beq E'\equiv \frac{mn}{\sqrt\Delta}\left(\frac{q}{k}c'd'\, Q^2
+\frac{k}{q}a'b'\, P^2 - (a'd'+b'c')Q\cdot P\right) \eeq

This curve intersects the real axis at the points:
\beq
\frac{p}{l}\frac{b'}{d'},\qquad \frac{p}{l}\frac{a'}{c'},\qquad
\eeq
For a fixed value of $\frac{p}{l}$, the set of intersection points is
in one to one correspondence with those for the primitive case, where
using Ford circles (or the methods of Ref.\cite{Sen:2007vb}) we saw
that curves of marginal stability do not intersect. However the value
of $\frac{p}{l}$ is not fixed. For given $m,n$ specifying a
non-primitive dyon, \eref{mnfact} permits several solutions for $p$
and $l$ in general. For each of them we obtain a construction in 1-1
correspondence with the set of curves of marginal stability for the
primitive case, and it appears quite likely that curves from one of
these sets can intersect with curves from another set. This would
result in curves of marginal stability that intersect each other in
the upper half plane.

To generate examples, it is convenient to revert to the notation in
which the charges of the decay products are labelled by a matrix
of integers $\pmat{m_1}{r_1}{s_1}{n_1}$ satisfying
Eqs.(\ref{firstdet}) and (\ref{seconddet}). From these two equations
we find that:
\beq
m_1 n+ n_1 m=mn
\eeq
from which we see that $m_1$ is a multiple of $m$. We write:
\beq
m_1 = m\alpha_1
\eeq
where $\alpha_1$ is another integer. The equations now yield the
following general form for the matrix:
\beq 
\pmat{m_1}{r_1}{s_1}{n_1} =
\pmat{m\alpha_1}{\frac{mn\,\alpha_1(1-\alpha_1)}{s_1}}{s_1}{n(1-\alpha_1)}
\eeq
The strategy is now to choose a value for $\alpha_1$ and then look for
the set of $s_1$ that divide $mn\,\alpha_1(1-\alpha_1)$. Finally to
ensure that we are dealing with a two-body decay, we must check that
there is no overall common factor in either of the matrices 
\beq
\pmat{m_1}{r_1}{s_1}{n_1},\quad \pmat{m-m_1}{-r_1}{-s_1}{n-n_1}
\eeq
In this way we can generate a large number of examples of curves of
marginal stability for non-primitive dyons decaying into a pair of
$\half$-BPS dyons. 

To check the possible intersections of such curves, we recall that
they intersect the real axis in the points $\frac{m_1-m}{s_1}$ and
$\frac{m_1}{s_1}$. If two such intervals intersect then the curves
will necessarily intersect in the upper half-plane.  Let us consider a
definite example. Suppose $(m,n)=(2,3)$. Then choosing $\alpha_1=1$,
we see that $s_1$ can be arbitrary. On the other hand choosing
$\alpha_1=2$ we find that the allowed values of $s_1$ are
$1,2,3,4,6,12$. It is easy to check that for the very simplest choices
the curves do not intersect. However, picking $\alpha_1=1,s_1=7$ and
$\alpha_1=2,s_1=12$ we find that all the conditions are satisfied and
the decay products are given by the matrices:
\bea
&\alpha_1=1,~s_1=7:&\qquad \pmat{2}{0}{7}{0},\quad
\pmat{0}{0}{-7}{3}\nn\\
&\alpha_1=2,~s_1=12:&\qquad \pmat{4}{-1}{12}{-3},\quad
\pmat{-2}{1}{-12}{6}
\eea
In terms of the integers $a,b,c,d$ the two decay processes are
parametrised by the matrices:
\bea
(i)\quad \pmat{a}{b}{c}{d}&=&\pmat{2}{0}{7}{1}\nn\\
(ii)\quad \pmat{a}{b}{c}{d}&=&\pmat{1}{1}{3}{4}
\eea
Each matrix satisfies $3ad-2bc=6$. 

Now the curves of marginal stability for the two decay modes
intersect the real axis at the following values:
\bea (i)\quad \tau_1 &=& 0,~\frac{2}{7}\nn\\[2mm] (ii)\quad \tau_1 &=&
\frac{1}{6},~\frac{1}{3} \eea These two intervals are overlapping,
hence the associated curves must intersect in the interior of the
upper half plane. We conclude that curves of marginal stability for
the decay of non-primitive dyons can in general intersect in the upper
half plane, unlike what happens for primitive dyons.  It would be
important to understand the physical and mathematical reasons why the
curves intersect, as well as the consequences of this fact.

\section{Analysis of the marginal stability curves: $\frac14$-BPS decay products}

\subsection{Decays into a $\half$-BPS and a $\frac14$-BPS dyon}

We now consider decays of a $\frac14$-BPS dyon into one $\half$-BPS
and one $\frac14$-BPS dyon. This is parametrised as in
\eref{gendecayR}. If the first decay product is taken to be
$\half$-BPS then we must impose the condition \eref{firstdet} which is
solved by \eref{firstcond}. However, the coefficients
$m_i,r_i,s_i,n_i$ are no longer required to be integers and therefore
nor are $a,b,c,d$. Moreover we want the second state to be
$\frac14$-BPS and therefore $adn-bcm\ne mn$. Finally, as indicated
earlier, we must check that the curve we obtain from \eref{cms}
actually describes the forward and not the reverse decay process.

Consider the case where the initial state is a primitive dyon with
$(m,n)=(1,1)$. For this case we find the curve of marginal stability
to be:
\beq
\label{halfquartercms}
\left(\tau_1-\frac{m_1- n_1}{2s_1}\right)^2
+ \left(\tau_2 + \frac{E}{2s_1}\right)^2
= \frac{1}{4s_1^2}\Big((m_1-n_1)^2 + 4r_1s_1 + E^2\Big)
\eeq
where
\beq
E\equiv \frac{1}{\sqrt\Delta}\left(s_1\, Q_R^2 -r_1\, P_R^2 - (m_1 - n_1)Q_R\cdot
P_R\right)
\eeq
On replacing $m_1,r_1,s_1,n_1$ by their expressions in terms of
$a,b,c,d$ we can also bring it to the form:
\beq
\label{halfquartercmsabcd}
\left(\tau_1-\frac{ad+bc}{2cd}\right)^2
+ \left(\tau_2 + \frac{E}{2cd}\right)^2
= \frac{1}{4c^2d^2}\left((ad-bc)^2 + E^2\right)
\eeq
with
\beq
E\equiv \frac{1}{\sqrt\Delta}\left(cd\, Q_R^2 +ab\, P_R^2 - (ad+bc)Q_R\cdot
P_R\right)
\eeq
The equation is very similar to the Sen circle for decays of a
primitive dyon into $\half$-BPS decay products. However, the
constraints on $a,b,c,d$ are quite different. Instead of analysing
this case further, we will return to it as a special case of the more
general decay into two $\frac14$-BPS states.

\subsection{Decays into two $\frac14$-BPS dyons}

We now address the case in which the initial $\frac14$-BPS dyon decays
into a pair of $\frac14$-BPS dyons. Again we start with the primitive
case, $(m,n)=(1,1)$. The relevant curve of marginal stability is the
same as in the previous subsection, \eref{halfquartercms}, except that
the determinants of $\pmat{m_i}{r_i}{s_i}{n_i}$ are both
nonzero. (Later we will also be able to specialise to the case where
one of them is zero.)

Let us now address the constraints on the final state parameters that
are required to ensure that the decay process corresponds to the
correct branch of \eref{marginalquartic}.  First of all, the quantity
$\Delta$ that appears in the BPS mass formula \eref{genbpsmass}
involves a square root, and we have taken all square roots to be
positive. This has the following consequence. Observe that:
\beq
\Delta(m_i\vQ+r_i\vP,s_i\vQ + n_i\vP)=
\det \pmat{m_i}{r_i}{s_i}{n_i}\Delta(\vQ,\vP)
\eeq
Positivity of $\Delta$ on both sides of the equation imposes the
condition:
\beq
\label{matpos}
\det \pmat{m_i}{r_i}{s_i}{n_i}> 0,\qquad i=1,2
\eeq
Since
\beq
\pmat{m_2}{r_2}{s_2}{n_2} = \pmat{1-m_1}{-r_1}{-s_1}{1-n_1}
\eeq
we find that:
\beq
m_1n_1-r_1s_1 > {\rm max}\,(m_1+n_1-1, 0)
\eeq
For what follows, it will be convenient to introduce the eigenvalues
$\beta_1,\gamma_1$ of $\pmat{m_1}{r_1}{s_1}{n_1}$ and
$\beta_2,\gamma_2$ of $\pmat{m_2}{r_2}{s_2}{n_2}$. Because the two
matrices commute (they are of the form $\bF$ and $1-\bF$) they can be
simultaneously diagonalised, from which we see that:
\beq
\beta_1 + \beta_2 = 1 = \gamma_1 + \gamma_2
\eeq
From the determinant conditions above, we have:
\beq
\beta_1\gamma_1>0,\quad (1-\beta_1)(1-\gamma_1)>0
\eeq

We will now examine the quantities $\frac{M_1}{M},\frac{M_2}{M}$ on
the curve \eref{halfquartercms}. For convenience, we would like to
choose a particular point on the curve and evaluate these quantities
there. The possible results are as follows.  If we find
$\frac{M_1}{M}>1$ at a point, then the marginal stability curve cannot
correspond to $M=M_1+M_2$. It may correspond to either $M_1=M+M_2$ or
$M_2=M+M_1$. Which of the two cases it corresponds to is then not very
important, but can be distinguished by looking at $\frac{M_2}{M}$. If
on the other hand we find $\frac{M_1}{M}<1$ then we have the
possibilities of being on the correct branch $M=M_1+M_2$ or on the
wrong branch $M_2=M+M_1$. This time it is essential to distinguish the
two, which can again be done by evaluating $\frac{M_2}{M}$. Being on
the correct branch requires $\frac{M_i}{M}<1$ for both $i=1$ and 2.

In any of these cases, having determined the relevant branch of
\eref{marginalquartic} at one point on the curve, we can be sure that
we will not cross over to another branch elsewhere on the same curve,
since crossing from one branch to another requires passing through a
point where one of the masses vanishes. But the BPS mass formula does
not vanish for any value of the moduli, so this is not possible
(unless the charges of that state vanish identically).

Let us assume that the matrix $\pmat{m_1}{r_1}{s_1}{n_1}$ is such that
the curve \eref{halfquartercms} intersects the real axis. This will
happen if the eigenvalues $\beta_1,\gamma_1$ are both real (without
loss of generality we take $\gamma_1\ge\beta_1$).  Then, a convenient
point at which to evaluate the mass ratios is one of the intersection
points of the curve with the real axis.  Setting $\tau_2=0$ in
\eref{cms}, we get the following equation for $\tau_1$:
\beq
\label{proportional}
n_1-\frac{r_1}{\tau_1} = m_1-\tau_1 s_1
\eeq
Now let us consider the expression $\frac{M_1^2}{M^2}$ in the
limit $\tau_2\ra 0$. We have:
\bea \nn \left.\frac{M_1^2}{M^2}\right|_{\tau_2\ra 0}&=&
\frac{[(m_1\vQ_R+r_1\vP_R)-\tau_1(s_1\vQ_R+n_1\vP_R)]^2}
     {[\vQ_R-\tau_1\vP_R]^2}\\
     &=&\frac{[(m_1-\tau_1s_1)\vQ_R-
\tau_1(n_1-\frac{r_1}{\tau_1})\vP_R]^2}{[\vQ_R-\tau_1\vP_R]^2}
     \eea
Using \eref{proportional} we now get:
\beq
\label{mratio}
 \left.\frac{M_1}{M}\right|_{\tau_2\ra 0}=|m_1-\tau_1s_1|
\eeq
On the real axis, \eref{halfquartercms} gives:
\beq
\tau_1 = \frac{1}{2s_1}\Big(\pm(\gamma_1-\beta_1)|+ (m_1-n_1)\Big)
\eeq
Inserting this into \eref{mratio} we find:
\beq
\label{finalratio}
m_1-\tau_1 s_1 = \beta_1~{\rm or}~\gamma_1
\eeq

Let us first consider the case $m_1n_1-r_1s_1 > 1$. We will show that
in this region the decay is not the desired one, but corresponds
instead to a branch of \eref{marginalquartic} describing a reverse
decay. With this condition on the determinant, one of the eigenvalues
(say $\gamma_1$) must be $>1$. Positivity of the second determinant,
which equals $(1-\beta_1)(1-\gamma_1)$, tells us that if $\gamma_1>1$
then also $\beta_1>1$. Thus we have that both eigenvalues are $>1$. It
follows that $\frac{M_1}{M}>1$ and we are, as promised, on the wrong
branch.

Next suppose $m_1n_1-r_1s_1 = 1$. The above considerations then show
that $\beta_1=\gamma_1=1$. Then we $\frac{M_1}{M}=1$. This means
$M_2=0$ and therefore the charges associated to the second state are
identically zero. In other words,
$\pmat{m_1}{r_1}{s_1}{n_1}=\pmat{0}{0}{0}{0}$. This is a trivial case
where the first decay product is the original state itself.

Let us note at this point that if $m_1,r_1,s_1,n_1$ had been taken to
be integers, and the corresponding state was restricted not to be
$\half$-BPS, we would necessarily have $m_1n_1-r_1s_1 \ge 1$. We have
shown that all such cases do not correspond to a valid decay of $M$
into $M_1$ and $M_2$, therefore no such decays exist for integer
coefficients. This is one of the key results of Ref.\cite{Sen:2007nz}.

That leaves the case 
\beq
\label{fraccase}
0<m_1n_1-r_1s_1 < 1,\quad 0<m_2n_2-r_2s_2 < 1
\eeq
which can only be satisfied for fractional coefficients. 

Requiring $\beta_1\gamma_1<1$ and also
$\beta_2\gamma_2=(1-\beta_1)(1-\gamma_1)<1$ we see that
$0<\beta_1,\gamma_1 <1$ and $0<\beta_2,\gamma_2 <1$. From this and
\eref{finalratio} we find $\frac{M_1}{M}<1,\frac{M_2}{M}<1$ and this
indeed corresponds to the decay process that we were looking for. Thus
\eref{fraccase} provides a necessary condition for the coefficients
$m_1,n_1,r_1,s_1$ in \eref{gendecayR} in order to have a decay of the
original dyon into two $\frac14$-BPS dyons. Under this condition, our
curve \eref{cms} describes the marginal stability locus in the
$\tau_1,\tau_2$ plane. However this is a locus of co-dimension 2 in
the full moduli space, for the following reason. Fractional
$m_1,r_1,s_1,n_1$ means that the decay process in terms of the
original integral charge vectors was into states living outside the
$\vQ,\vP$ plane. This is precisely the case, referred to earlier,
where the moduli in $M$ need to be adjusted to make the final state
charges (after R projection) lie in the same plane as the initial
state charges\cite{Sen:2007nz}.  It remains to find a sufficient
condition on the values of $m_1,r_1,s_1,n_1$ as well as to understand
more precisely the condition on the moduli matrix $M$ which put the
projected charge vectors in the plane of the decaying dyon.

\section{Discussion}

We have found a general equation for marginal stability of
$\frac14$-BPS dyons to decay into two final state particles,
\eref{cms}. Analysis of the equation reveals many distinct cases with
different properties. We believe this analysis can be easily extended
to multi-particle final states. The construction of Ford circles and
especially their dual circles proved useful in this analysis and we
suspect that there may be a deeper mathematical relationship to the
Sen circles of marginal instability for primitive dyons decaying into
$\half$-BPS final states.

To complete the discussion of the previous section of decays into
$\frac14$-BPS states we need to consider the case where the
eigenvalues of $\pmat{m_1}{r_1}{s_1}{n_1}$ are complex, in which case
the curve of marginal stability does not intersect the real axis. Also
we need to generalise to the case of non-primitive initial states.

We see that decays into $\half$-BPS final states are labelled by
integers $a,b,c,d$ that are in $PSL(2,Z)$ for primitive initial states
and obey the more complicated relation \eref{secondcond} when the
initial state is a non-primitive dyon. However, such integers do not
occur when the final state consists of $\frac14$-BPS dyons. It would
be nice to understand the physical origin of this $PSL(2,Z)$ and its
generalisations in the cases where they occur. We expect this can be
done through the string
network\cite{Aharony:1997bh,Dasgupta:1997pu,Sen:1997xi} representation
of the dyons.

The role of non-primitive dyons and their decays has been somewhat
mysterious\footnote{and needs to be taken up by the Department of
  Mysteries, Ministry of Magic.} since the observation of
Ref.\cite{Dabholkar:2007vk} that the quantum states of such dyons are
not counted by the famous genus-2 modular form of
Ref.\cite{Dijkgraaf:1996it} but appear to be connected to a
higher-genus Riemann surface. We have exhibited the $\half$-BPS decays
of such dyons and noted that their curves of marginal stability
intersect in the interior of moduli space. This may be helpful in
resolving the puzzle of their role in the counting problem.

\section*{Acknowledgements}

We are grateful to Suresh Nampuri, Nitin Nitsure and Ashoke Sen for
helpful discussions. We thank the people of India for generously
supporting our research. The work of A.M. was supported in part by
CSIR Award No. 9/9/256(SPM-5)/2K2/EMR-I. S.M. would like to
acknowledge the generous hospitality of the Pacific Institute of
Theoretical Physics, UBC Vancouver, and of the Galileo Galilei
Institute, Firenze, where parts of this work were done.

\bibliographystyle{JHEP}

\bibliography{14bpsdecay}

\end{document}